\documentclass[conference]{IEEEtran}

\usepackage{graphicx}
\usepackage{wrapfig}

\usepackage{lipsum}
\usepackage{dirtytalk}
\usepackage[utf8]{inputenc}
\usepackage{subfigure}
\usepackage{mathtools}
\usepackage[algo2e,linesnumbered]{algorithm2e} 
\usepackage{algorithmic}
\usepackage{algorithm}
\usepackage{dirtytalk}
\usepackage[utf8]{inputenc}
\usepackage[english]{babel}
\usepackage{amsthm}
\usepackage{amsmath}
\usepackage{lipsum}
\newcommand{\subparagraph}{}
\usepackage{titlesec}

\usepackage[colorlinks]{hyperref}

\theoremstyle{definition}

\makeatletter
\newcommand*{\rom}[1]{\expandafter\@slowromancap\romannumeral #1@}
\makeatother

\showoutput
\titlespacing\section{0pt}{5pt plus 2pt minus 2pt}{5pt plus 2pt minus 2pt}
\titlespacing\subsection{0pt}{3pt plus 2pt minus 2pt}{3pt plus 2pt minus 2pt}
\titlespacing\subsubsection{0pt}{2pt plus 2pt minus 2pt}{2pt plus 2pt minus 2pt}
\usepackage{bbding}

\begin{document}
\long\def\/*#1*/{}
\setlength{\abovedisplayskip}{1pt}
\setlength{\belowdisplayskip}{1pt}
\title{Energy Aware Competitiveness Power Control in Relay-Assisted Interference Body Networks}

\author{\IEEEauthorblockN{Mohamad Ali\IEEEauthorrefmark{1}, Hassine Moungla\IEEEauthorrefmark{2}\IEEEauthorrefmark{1}, Ahmed Mehaoua\IEEEauthorrefmark{1}}
\IEEEauthorblockA{\IEEEauthorrefmark{1}LIPADE, University of Paris Descartes, Sorbonne Paris Cit\'{e}, Paris, France}{\IEEEauthorrefmark{2}UMR 5157, CNRS, Institute Mines-Telecom, T\'{e}l\'{e}com SudParis, Nano-Innov CEA Saclay, France}\\Email: \{mohamad.ali; hassine.moungla; ahmed.mehaoua\}@parisdescartes.fr}

\maketitle
\begin{abstract}
Recent advances in microelectronics have enabled the realization of Wireless Body Area Networks (WBANs). Increasing the transmission power of WBAN's nodes improves the Signal to Interference plus Noise Ratio (SINR), and hence decreases the bit error probability. However, this increase may impose interference on nodes within the same WBAN or on other nodes of nearby coexisting WBANs, as these WBANs may use similar frequencies. Due to co-channel interference, packet collisions and retransmissions are increased and consequently, the power consumption of the individual WBANs may increase correspondingly. To address this problem, we adopt the approach of two-hop cooperative communication due to its efficiency in power savings. In this paper, we propose a cooperative power control-based algorithm, namely, IMA, for interference mitigation among the individual sensors of a single WBAN. Basically, our approach selects an optimal set of relays from the nodes within each WBAN to mitigate the interference. Thus, IMA selection criterion relies on the best channel, namely, SINR and power conditions to select the set of best relays. The experimental results illustrate that IMA improves the SINR, the power efficiency and extends WBAN lifetime. In addition, the results illustrate that IMA lowers the bit error probability and improves the throughput.
\end{abstract}

\section{Introduction}
The pervasive use of wireless networks, recent developments in micro-electronics and the miniaturization of low-power sensors led to the existence of WBANs. The technological advancements in the wireless communication enabled low-power, intelligent, miniaturized sensor nodes placed in or around the human body to enhance the quality of patient's life. Various applications such as personal health monitoring, ubiquitous healthcare, sports, entertainment, and military have been found to provide high reliable data communication systems as well as to improve the healthcare conditions. Their individual sensors mainly monitor physical activities, actions and vital signs as glucose percentage in blood, heart beats, respiration, body temperature and/or can record electrocardiography (ECG) \cite{key2,key9,key14}.

The co-channel interference is challenging due to the highly mobile and resource constrained nature of WBANs. Firstly, such nature makes the allocation of a global coordinator to manage the coexistence problem among coexisting WBANs as well as the application of advanced antenna and power control techniques used in other networks unsuitable for WBANs. Secondly, the stringent resource in WBANs is the energy, which requires a careful calibration of the transmission power of the individual nodes to minimize interference in the network. Thirdly, due to the infeasibility of coordination among the WBANs and the unpredictable movements of their individual sensors, an interference may arise due to the simultaneous transmissions of sensors of different coexisting WBANs using the same channel. For instance, an interference may happen when the simultaneous reception from two or more distinct sensors of different WBANs collide at the same receiving node such as a WBAN's coordinator. Consequently, the interference may affect the communication links and degrade the performance of each individual WBAN \cite{key15}. Therefore, interference mitigation is of the utmost importance to improve the reliability of the whole network. To this end, the IEEE standard proposes three mechanisms for co-channel interference mitigation in WBANs, namely, beacon shifting, channel hopping and active superframe interleaving \cite{key26}. In this paper, we tackle these issues and contribute the following:
\begin{itemize}
 \item  \textit{IMA, a cooperative power control-based scheme for interference mitigation among the individual sensors of a single WBAN}
 \item \textit{A relay selection criterion that determines an optimal set of relays using the best-known channel and node energy conditions}
\end{itemize}
The simulation results show that our proposed approach can significantly lower the interference among the individual sensors of an intra-WBAN as well as increase the power savings at both node- and WBAN- levels. Moreover, IMA significantly avoids the intra-WBAN interference and does not require any mutual coordination among the individual WBAN coordinators. The rest of the paper is organized as follows. Section \rom{2} sets our work apart from other approaches in the literature. Section \rom{3} summarizes the system model and the assumptions. Section \rom{4} describes \textit{IMA} in detail. Section \rom{5} presents the simulation results. Finally, the paper is concluded in Section \rom{6}

\section{Related Work}
The problem of intra-WBAN interference has been addressed through cooperation, power control, link adaption, multiple access and resource allocation schemes. Example schemes that pursued the cooperation methodology include \cite{key7}, \cite{key22}, \cite{key17}. Dong et al., \cite{key7} proposed a single-relay cooperative scheme where the best relays are eventually selected in a distributed fashion. The scheme is based on MAC request-to-send (RTS) and clear-to-send (CTS) signaling, where a set of potential relays compute individually the required transmission power to participate in the cooperative communication to significantly mitigate the interference. Also, Dong et al., \cite{key22} addressed the problem of coexistence of multiple non-coordinated WBANs. A decode-and-forward protocol with two relays and selection combining at the desired coordinator is used, enabling intra-network and inter-network operation, to allocate slots for each link packet transmission to mitigate intra-WBAN interference. Whilst, Feng et al., \cite{key17} proposed a prediction-based dynamic relay transmission scheme through which the problem of "when to relay" and "who to relay" is decided in an optimal way. Other approaches pursued power control schemes for intra-WBAN interference mitigation. Dong et al., \cite{key9} proposed a joint two-hop relay-assisted cooperative communication integrated with transmit power control for intra-WBAN interference mitigation. This scheme reduces co-channel interference at the WBAN individual nodes and extends the WBAN energy lifetime.

A number of approaches adopted medium access schemes to mitigate intra-WBAN interference include \cite{key6}, \cite{key3}. Mahapatro et al., \cite{key6} proposed a TDMA-based scheme that enables two or three coexisting WBANs to agree on a common TDMA schedule to reduce the intra-WBAN interference and the transmission latency per each individual node. Whereas, Chen et al., \cite{key3} proposed a distributed scheme that adopts a polling-based TDMA for traffic coordination of the individual sensors within a WBAN and a carrier sensing mechanism conducted before each beacon transmission to deal with inter-WBAN interference. Meantime, some approaches adopted the link adaption methodology to mitigate the intra-WBAN interference include \cite{key1}, \cite{key5}, \cite{key1}. Yang et al., \cite{key1} focused on the performance at the WBAN coordinator which periodically calculates SINR. Based on this calculation, it commands the nodes within its WBAN to select an appropriate scheme (data rate, modulation, duty cycles, etc.) to mitigate the interference. Meantime, Martelli et al., \cite{key5} considered a WBAN where the coordinator periodically queries its sensors to transmit data. The network adopts the carrier sense multiple access with collision avoidance (CSMA/CA) and the nodes adopt link adaptation to select the modulation scheme according to the experienced level of interference. Whilst, Domenicali et al., \cite{key30} analyzed the performance of a reference WBAN in terms of bit error rate, throughput, and energy lifetime. The study proved the performance of a WBAN can be improved by the adoption of an optimized time hopping code assignment strategy. A strategy to extend the lifetime of the WBAN is also introduced.

Jamthe et al., \cite{key23} pursued the medium access methodology and proposed a quality of service based medium access control (MAC) scheduling approach to avoid inter-WBAN interference and introduce a fuzzy inference engine for intra-WBAN scheduling so as to avoid interference within WBANs. Other approaches pursued the resource allocation include \cite{key20}, \cite{key16}. Liang  et al., \cite{key20} proposed a distributed interference detection and mitigation scheme through using adaptive channel hopping for intra- and inter-WBAN interference mitigation. Whilst, Movassaghi et al., \cite{key16} proposed a dynamic resource allocation scheme for intra- and inter- WBAN interference mitigation among multiple coexisting WBANs through using orthogonal channels for high interfering nodes. 

Though, most of the recent works addressed problems related to co-channel interference at the WBAN-level and do not consider the node-level interference. In this paper, we take a step forward, and exploit the benefits of two-hop cooperation, and propose a cooperative power control-based algorithm for interference mitigation among the individual sensors of a single WBAN. Thus, we depend on the relaying and two-hop communication scheme to minimize intra-WBAN interference and increase the power savings at the node- and WBAN-levels.

\section{System Model and Assumptions}
We consider the realistic scenario when multiple beacon-enabled CSMA/CA-based WBANs coexist in a large hall of a hospital. Each WBAN consists of a single coordinator and up to K sensors, each generates its data based on a predefined sampling rate and transmits data at a maximum rate of 250Kb/s within the license-free 2.4 GHz band. Unlike sensors, we assume all coordinators are equipped with unconstrained energy supply. In fact, the time period between any two consecutive beacons depends on the application, the required bit rate, and the transmitted packet size. The smaller the period is, the smaller data packet size is, the more frequent contention for the channel and the smaller the throughput is. Thus, we assume the followings:
\begin{itemize}
  \item A WBAN topology based on two-hop communication
  \item A CSMA/CA as a medium access scheme
 \end{itemize}
\section{Intra-WBAN Interference Mitigation Approach - IMA}
As pointed out, a co-channel interference may arise due to the collisions amongst the concurrent transmissions made by sensors in different WBANs using the same channel. To address this issue, we exploit the benefits of the two-hop relaying schemes in order to tackle this problem and reduce the probability of collision while enabling autonomous scheduling of the medium access within each WBAN. Therefore, IMA minimizes the intra-WBAN interference and maximizes the energy lifetime. The coordinator of each WBAN periodically broadcasts beacons, through which the nodes synchronize their transmissions according to its clock. Then, after the successful reception of each beacon, the nodes within each WBAN compete for the channel using CSMA/CA. During the contention access period (CAP), each WBAN's node pursues the communication model with its associated coordinator as follows, RTS-CTS-DATA-Acknowledgment (ACK). Received beacons and RTS packets are used in SINR computation and remain valid as long as they happen to be within the coherence time of the channel. IMA involves four phases; Beacon phase, RTS phase, CTS phase and $CTS_{Reception}$ phase.

\subsection{Beacon phase}
In this phase, the coordinator periodically broadcasts beacons every T milliseconds (ms) to enable the sensors within its WBAN to synchronize. Based on the power contained within the beacon signal, each node computes the corresponding SINR which is defined at a given node as the power received from the desired transmitter divided by the sum of undesired powers received from other interfering nodes plus the power contained within the noise. Basically, it is used to evaluate the channel quality along with any path from a source to the coordinator through relays. Thus, R denotes the set of WBAN's nodes, where each node can successfully decode the beacon and whose \textit{$SINR >= SINR_{Thr}$}. SINR is computed as in \textbf{eq.(1)}. Where P is the desired power at receiver, $I_{i}$ is the interference power received from interfering node \textit{i} and $N_{0}$ is an additive white Gaussian noise 
\begin{equation}
SINR =\cfrac{P}{\sum_{i=1}^{N} I_i + N_0}
\end{equation}
\subsection{RTS phase}
After the successful reception of a beacon, All nodes that have pending data (denoted by sources) start a contention for the channel, where each node transmits its short packet ($RTS_{i}$). The transmissions of such RTS packets may introduce interference at some other nodes. Each node that is in the reception mode when sensing these RTS packets, it immediately computes the SINR of the received signal. If SINR is above a threshold, i.e., \textit{$SINR >= SINR_{Thr}$}, this node will be included in the relay candidate set. Thus, each relay candidate has successfully received a beacon and RTS packet can compete for the relay selection process in the next phase. Therefore, the periodic computations evaluate the channel quality from each relay candidate to its corresponding source node. Up to this information, each relay candidate can evaluate the communication path from a particular source node to the coordinator through itself. We define:
\begin{itemize}
  \item S denotes the set of all source nodes have pending data to transmit in the current superframe
  \item $N_{i}$ denotes the set of neighbors of a source node denoted by $S_{i}$ that successfully decode $RTS_{i}$
\end{itemize}
\subsection{CTS phase}
Each relay candidate has the knowledge of the channel condition from itself to the source nodes and to the coordinator. Basically, all nodes that have low SINR values are excluded from the relay selection process to avoid the interference. The main goal is to select a relay for each source node so that the interference is minimized and the power saving is maximized. This way, each relay sets a timer to compute a waiting time value. Then, that relay computes the SNR and SINR values from the signals received from the coordinator and the source node, respectively. Consequently, the smaller the difference between both values (SNR and SINR) at each relay implies the better the candidate is likely to be chosen as the best relay. knowing that the waiting time is inversely proportional to their values difference. This difference follows a threshold margin, hence, each relay candidate within the set $Q_{i}$ of $S_{i}$ transmits its CTS to $S_{i}$, where $Q_{i}$ denotes the set of nodes successfully decode the $RTS_{i}$ and the beacon, $Q_{i} = R \cap N_{i}$. Then we define the beacon validity as:

\textbf{if (reception time of RTS - reception time of beacon) $<$ T ms, then it is valid, otherwise, it is not}
\subsection{$CTS_{RECEPTION}$ phase}
Upon the reception of CTS packets from the relay candidates, each source node waits a period of time and enqueues the last two CTS packets received from the last two distinct relays. Then, it selects the best relay based on the high SINR and residual energy (ER) of the battery reported within each CTS packet. M denotes the set of nodes successfully decode the CTS.
\textbf{Algorithm \ref{algo_ima}} provides a summary of IMA.
\IncMargin{1em}
\begin{algorithm}
\SetKwData{Left}{left}\SetKwData{This}{this}\SetKwData{Up}{up}
\SetKwFunction{Union}{Union}\SetKwFunction{FindCompress}{FindCompress}
\SetKwInOut{Input}{input}\SetKwInOut{Output}{output}

\Output{Optimal set of relays}
 
 Coordinator C broadcasts beacon
 
\While{R isNotEmpty} { 
 
  $r_{i}$ computes $SNR_{i}$

 }
 
\While{true} {  
 
\While {S isNotEmpty}{

 $S_{i}$ broadcasts $RTS_{i}$ to  $N_{i}$ $\forall$  i=1, 2...s

\While{$N_{i}$ isNotEmpty} {

$n_{ij}$ computes $SINR_{ij}$ $\forall$ i, j

 }

 }

 }

 \While {true} {
 
  $S_{i}$ sets $Q_{i} =  R \cap N_{i}, \forall$ i = 1,2,...s

 \While{$Q_{i}$ isNotEmpty} { 

 $n_{ik} \in Q_{i}$ computes $SNR_{ik}$

 $n_{ik} \in Q_{i}$, sets $Diff_{ik} = SNR_{ik} - SNR_{ik=j}$ $\forall k \leq sizeof(Q_{i})$

\If {$Diff_{ik} <= threshold$} {  

$Q_{i} = Q_{i} - n_{ik}$

 }
 
\Else 
{

 $n_{ik}$ waits  $t_{ik}= Rand_{t_{ik}} + Constant / f(Diff_{ik})$

 \If{isFreeChannel}  {
 
 $n_{ik}$ broadcasts $CTS_{ik}$

  }
 
\Else 
{
 
 \While{$CTS_{Retries} < Max_{Retries}$}  { 
 
 $n_{ik}$ backoffs then retries

 } 
 
 }
 
 }
 
 } 
 
 }
 
  \While{M isNotEmpty}
  {  

 $\forall m_{i} , i = 1, 2,\dots$
 
  \While{$Queue_{S_{i}}$ isNotFull}
  {
                                  
  $S_{i}$ waits $CTS_{ik}$  $\forall$ k
            
   \If{$m_{i}$ isReceived} {

   $m_{i}$ extracts $bat_{Residual_{ip}}$ $\&$ enqueues $n_{ik}$ 

  } 
  
  }
                             
   \If{$Queue_{S_{i}}$ isFull} {
                           
 $m_{i}$ sets $q_{ih} = MAX(bat_{Residual_{ip}}, Bat_{Residual_{ip+1}})$

 $S_{i}$ broadcasts $q_{ih}$ as  relay winner
        
 } 
 
 }

\caption{Interference Mitigation Algorithm - IMA}
\label{algo_ima}
\end{algorithm}
\DecMargin{1em}

\section{Performance Evaluation}
\subsection{Simulation Environment and Setup}
The simulation environment is based on the IEEE CSMA/CA specifications, where a small scale fading channel model and a channel coherence time of 500ms are considered for on-body communication \cite{key26}. The WBAN is located within a space of 3x3x3$m^{3}$ and consists of a coordinator and up to M=8 sensor nodes. We also used the industrial, scientific and medical (ISM) radio bands for operating frequency of 2.4 GHz. We compared the performance of a WBAN that employs IMA coexisting with other WBANs that do not employ IMA. To have further fair results and to illustrate the outperformance of our algorithm, we generate the same simulation setup in terms of topology, the channel model, the medium access control (MAC) and the physical (PHY) specifications which involve the node transmission power (-10 dBm), receiver sensitivity (-84.7 dBm), noise floor (-102 dBm), an additive white Gaussian noise (AWGN) of mean zero and $\sigma$=6.81 dBm as well as path loss exponent (4.22) \cite{key26}. In this work, we focus on three performance metrics; SINR, lifetime and throughput, which have the direct impact on the performance of each individual WBAN.
\begin{enumerate}
\item \textbf{WBAN Lifetime} is defined as the sum of residual energies of WBAN nodes at a particular point in time.
\item \textbf{Battery energy residue} is defined as the amount of energy stored within a particular node's battery at a particular point in time.
\item \textbf{Throughput} is defined as the sum of the number of successful packets delivered in a period of time at a particular node.
\item \textbf{Outage probability (OP)} is defined as the probability of SINR being below a given threshold.
\end{enumerate}
\begin{equation} 
OP = Pr(SINR \leq SINR_{threshold})
\end{equation}
\begin{figure}
  %\centering
    \begin{center}
    \includegraphics[width=0.35\textwidth, height=0.2\textheight]{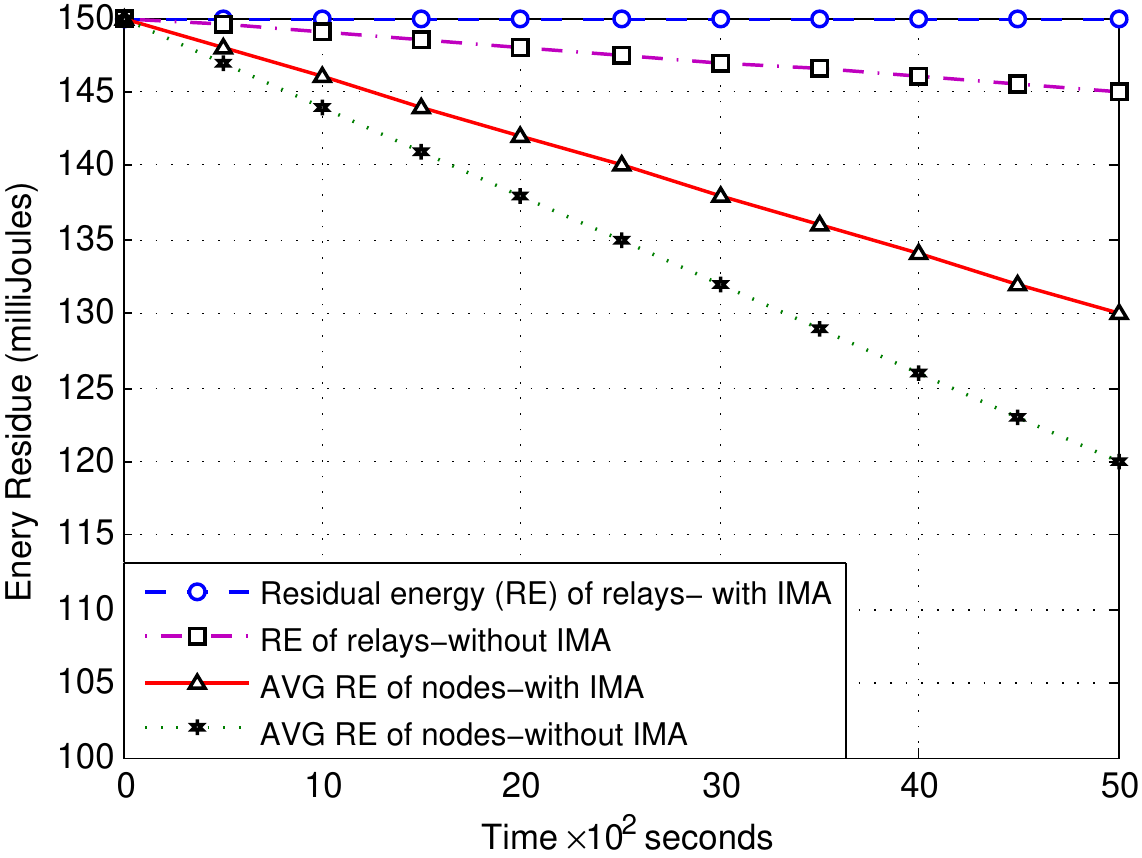}
      \end{center}
     \caption{Battery residual energy of coordinator, relays and ordinary and sensor nodes }
      \label{fig:nodeenergy}
\end{figure}
 \subsection{Network lifetime and battery energy residue}
 All the nodes of a WBAN have the same initial battery energy (150 milliJoules). The energy residue of IMA scheme denoted by RE and that of the competing scheme, namely, without-IM, are compared in \textbf{Figure \ref{fig:nodeenergy}}. As can be seen from the figure, RE of IMA is always higher than that of the competing scheme all the time. Referring to curves in the figure, it is evident to notice that RE values of both schemes decrease linearly with time. The AVGRE results of with-IMA are always larger than the AVGRE results of without-IMA. with-IMA's RE starts at 150 milliJoules and eventually stabilizes at 150. However, without-IMA's RE decreases very slightly to eventually stabilize to 145 milliJoules. Whilst, in without-IMA, the AVGRE starts at 150 milliJoules and decreases faster than with-IMA to stabilize at 130 milliJoules, while the second stabilizes at 120. The improvement in energy savings and the extension of WBAN lifetime is due to some reasons. 1) IMA selects relay nodes of minimal interferences that were supposed to share in the RTS/CTS contention phase. Such relays are now avoided with the employment of IMA algorithm and so they do not transmit their CTS packets accordingly. Moreover, these relays do not receive their respective ACKs as well as do not introduce interferences at other nodes. Consequently, the number of communicated packets in the system as well as the probability of collisions and hence energy consumption. Also, with IMA employment, no need for nodes that receive only RTS to reply with their CTS which decreases packet transmissions are reduced which lowers the energy consumption. 2) IMA selects nodes with higher SINR values which improve the energy per bit, which increases the probability of a successful delivery of packets at some nodes and hence decreases the number of retries for retransmissions.
 \subsection{Throughput}
\begin{figure}
  \centering
        \includegraphics[width=0.35\textwidth, height=0.2\textheight]{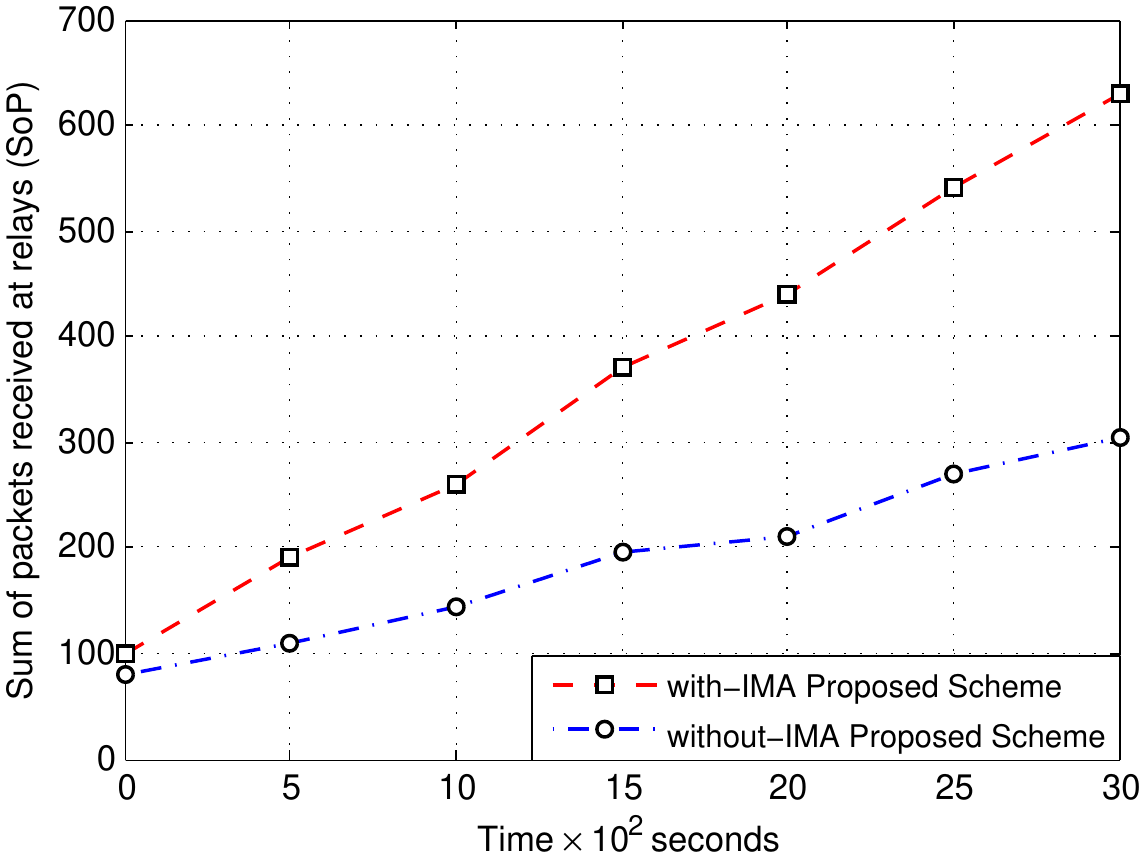}
\caption{Throughput results with and without employment of IMA}
  \label{fig:bb}
\end{figure}

The sum of successful packets received (SoP) at the WBAN coordinator versus time for a WBAN that employs IMA and another that does not employ IMA are compared in \textbf{Figure \ref{fig:bb}}. As can be clearly seen in the figure, IMA always provides a higher SoP than the competing scheme (another scheme that does not employ IMA) all the time. Referring to the curves in the figure, for instance, at time=0, both results have similar SoP values and hence, as time increases, SoP of both schemes increase linearly in different speeds, i.e, the curve of IMA is faster than that of the another scheme. The difference between their SoPs curves becomes evident at time=3000 seconds, SoP=300 packets successfully received at the WBAN coordinator that does not employ IMA, whereas, SoP=630 packets received at the WBAN coordinator that employs IMA. The improvement of the throughput in IMA is due to the increase in the SINR and the avoidance of interfering nodes. Hence, the achievement of higher SINR increases the energy per bit; the signal becomes less error-prone and more resistive to interferences, also the fraction of bit errors is decreased and which decreases the packet retransmissions as well. Furthermore, the throughput is increased due to the decrease in the number of collisions and packet retransmissions. In addition, IMA algorithm selects best relays with minimal interference level which increases the chances for successful packets delivery.
\subsection{Discussion of RTS and CTS}
IMA suggests that any node receives both beacon and RTS  packets with a precise SINR difference margin contends for the relay selection process. Firstly, these nodes that receive RTS packets only are excluded from the selection process and hence can not reply by their CTS packets which decrease the number of CTS packets. Secondly, IMA avoids nodes that experience high interference (with minimal SINR) even though they overhear both the coordinator (beacons) and the source nodes (RTS packets). These nodes are also excluded from the selection process and hence they do not transmit their CTS packets. Consequently, this lowers the number of nodes compete for the selection process as well as decreases the number of exchanged CTS and their corresponding ACKs. As a result, the total number of RTS/CTS and ACKs are decreased in the WBAN.
\section{Conclusion}
In this paper, we have presented \textit{IMA}, a distributed CSMA-based power control-based for intra-WBAN interference mitigation scheme based on two-hop cooperative communication. \textit{IMA} exploits the benefits of the two-hop scheme to lower the probability of collisions among transmission of sensors within a single WBAN as well as among the different coexisting \textit{WBAN}s. Accordingly, each distinct sensor within the WBAN selects the best relay to retransmits its packet to the coordinator. Compared with other sample algorithm, \textit{IMA} has low complexity and does not require any inter-\textit{WBAN} coordination. Simulation results show that \textit{IMA} outperforms other sample schemes in terms of interference mitigation, energy savings, and throughput.

\end{document}